\documentclass{article}
\usepackage{spconf,amsmath,graphicx,hyperref}

\title{Silent Speech Sentence Recognition with Six-Axis Accelerometers using Conformer and CTC Algorithm}
\name{Yudong Xie \qquad Zhifeng Han \qquad Qinfan Xiao \qquad Liwei Liang \qquad Lu-Qi Tao$^{\star}$ \qquad Tian-Ling Ren$^{\star}$
\thanks{$^{\star}$Corresponding authors. This work is supported by the National Key R\&D Program
(2022YFB3204100, 2021YFC3002200), the National Natural Science
Foundation (U20A20168) of China, the Research Fund from Tsinghua
University Initiative Scientific Research Program, and a grant from the
Guoqiang Institute, Tsinghua University.}}
\address{School of Integrated Circuits and Beijing National Research Center for Information Science \\ and Technology, Tsinghua University, Beijing 100084, China}
\begin{document}
%
\maketitle
\begin{abstract}
Silent speech interfaces (SSI) are being actively developed to assist individuals with communication impairments who have long suffered from a reduced quality of life.  However, silent sentences are difficult to segment and recognize due to elision and linking. A novel silent speech sentence recognition method is proposed to convert the facial motion signals collected by six-axis accelerometers into transcribed words and sentences. A Conformer-based neural network with the Connectionist-Temporal-Classification algorithm gains contextual understanding and translates the non-acoustic signals into words sequences. Test results show that the proposed method achieves a 97.17\% accuracy in sentence recognition, surpassing the existing silent speech recognition methods with a typical accuracy of 85\%-95\%, and demonstrating the potential of accelerometers as an available SSI modality for high-accuracy silent speech sentence recognition.
\end{abstract}
\section{Introduction}
\label{sec:intro}

Speech has been one of the most essential ways for humans to communicate for thousands of years. Acoustic signals convey much of the important information in our daily lives. However, speech-based communication cannot be performed well under certain physiological constraints, especially for people with vocal defects, such as laryngectomy patients. For the voiceless population, the inability to communicate effectively can lead to profound social isolation and psychological distress. In emergency situations, the lack of a reliable communication method can even be life-threatening, as they are unable to call for help or express urgent needs. Therefore, the development of a non-invasive, portable, and highly accurate silent speech recognition system \cite{lee2021biosignal, schultz2017biosignal} is not just a technological advancement but a necessity for improving the quality of life and safety of these individuals.

Silent speech interfaces (SSI) have been explored with various modalities such as electromagnetic, mechanical, ultrasonic, and visual approaches \cite{denby2010silent, gonzalez2020silent, zheng2023speech, yang2023mixed}. However, these systems are often bulky, expensive, or impractical for daily use. Accelerometers offer a non-invasive, noise-robust, and portable alternative \cite{varanis2018mems, farrahi2019calibration}, making them suitable for continuous silent speech recognition.

Most silent speech recognition systems focus solely on word classification \cite{kwon2023novel, kadiri2024investigation, shafiq2024fusion, kwon2024speech}, with limited attention to continuous sentence recognition, which involves translating unvoiced sentences into word sequences from a predefined vocabulary. Sentence recognition remains challenging due to elision (loss of syllables) and linking (word boundaries blending) \cite{wang2009contributions}. None of the prior studies has effectively addressed the challenges of silent speech sentence recognition in dynamic and accurate segmentation and recognition \cite{wang2012sentence}. While some researchers reported high accuracy in sentence recognition, their approaches treated entire sentences as single entities, just like word-level classification. The nuanced segmentation and dynamic contextual understanding required for accurate sentence-level recognition were overlooked \cite{wang2012sentence, luo2021end}.

This paper introduces a novel silent speech interface (SSI) approach that captures facial motion signals using a six-axis accelerometer and decodes them dynamically using a Conformer-based architecture \cite{gulati2020conformer} with connectionist temporal classification (CTC) \cite{graves2014towards}. Unlike static whole-sentence classification, the system can learn dynamically from context to obtain high accuracy in segmenting and recognizing complex combinations of words and sentences. To the best of our knowledge, no prior study has reported the SSI method to segment and recognize sentences dynamically.

\section{Method}
\label{sec:method}

\subsection{Data Recording}

\subsubsection{Microelectromechanical System}

\begin{figure}[t]
\centering
\centerline{\includegraphics[width=8.5cm]{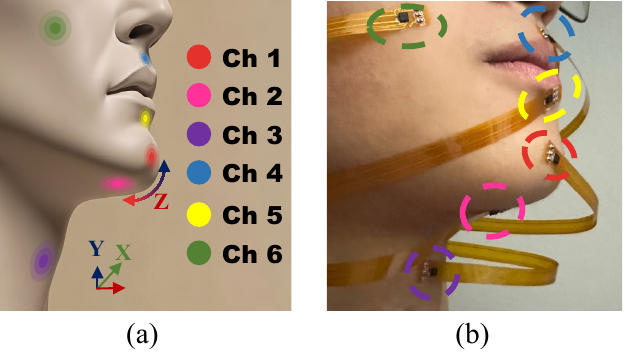}}
\caption{Location of the accelerometers used to collect facial motion signals: (a) Location of sensors; (b) A patient wearing the sensor system.}
\label{fig:location}
\end{figure}

The test was conducted using six accelerometers (channels). Each accelerometer records three-axis acceleration data and three-axis gyro data at the sampling frequency of 50 Hz. Fig.\ref{fig:location} shows the location of the accelerometers to record the facial motion data, and the designed wearable sensing system.

\subsubsection{Experiment}

One laryngect and three undergraduates participated in the test to evaluate the performance of the proposed method.

In total, sixteen English words, two English sentences, and eight Chinese phrases that are widely used in daily lives were recorded. Each word was recorded 100 times repeatedly, and each sample contained 80 sampling points, so the size of one sample is 80 (window size) × 6 (channels) × 6 (axes). As for a sentence or phrase, the window size is 180 and is repeated 30 times in the experiment. All the tested words and sentences as well as their meanings are listed in Table \ref{table_1}.

\begin{table}[b]
\caption{List of Words, Sentences, and Phrases}
\label{table_1}
\setlength{\tabcolsep}{3pt}
\centering
\renewcommand{\arraystretch}{1.5}
\begin{tabular}{|p{40pt}|p{200pt}|}
\hline

Words & Afternoon, Thanks, Beautiful, Wait, Breakfast, Want, Drink, Water, Hello, Welcome, Please, What, Sorry, Wonder, Test, Wonderful \\

\hline

Sentences & Drink water.

Hello, please wait. \\
\hline

Chinese Phrases & Tengtong (Hurt), Fanshen (Turn over), 

Xiachuang (Getting out of bed), Henkaixin (Very happy), 

Xiexieni (Thank you), Woyaoheshui (I want to drink water), 

Woxiangchifan (I want to have a meal), 

Tianqizhenhao (What nice weather)\\

\hline
\end{tabular}
\end{table}

The dataset is split 70:15:15 into training, validation, and test sets. The training set is augmented tenfold via word concatenation (into potentially meaningless "sentences") and by adding Gaussian noise with a standard deviation of one-third of the original values.

\subsection{Data Preprocessing}

The data were preprocessed using moving average (3-point smoothing), a 4th-order Butterworth high-pass filter at the 2Hz cutoff frequency to remove motion artifacts, and z-score normalization. Each sample was flattened to a final dimension of 36 $\times$ window size (80 for words, 180 for sentences) to improve data quality and model generalization.

\subsection{Machine Learning Model}

\subsubsection{Network Structure}

Sentence recognition in silent speech interfaces (SSI) is challenging due to the complexities such as elision and linking, especially in English. The complexities often make the traditional word-by-word approach ineffective. This study addresses this challenge with an end-to-end neural network based on Conformer with a multi-head local self-attention mechanism and CTC loss. 

\begin{figure}[t]
\centering
\centerline{\includegraphics[width=8.5cm]{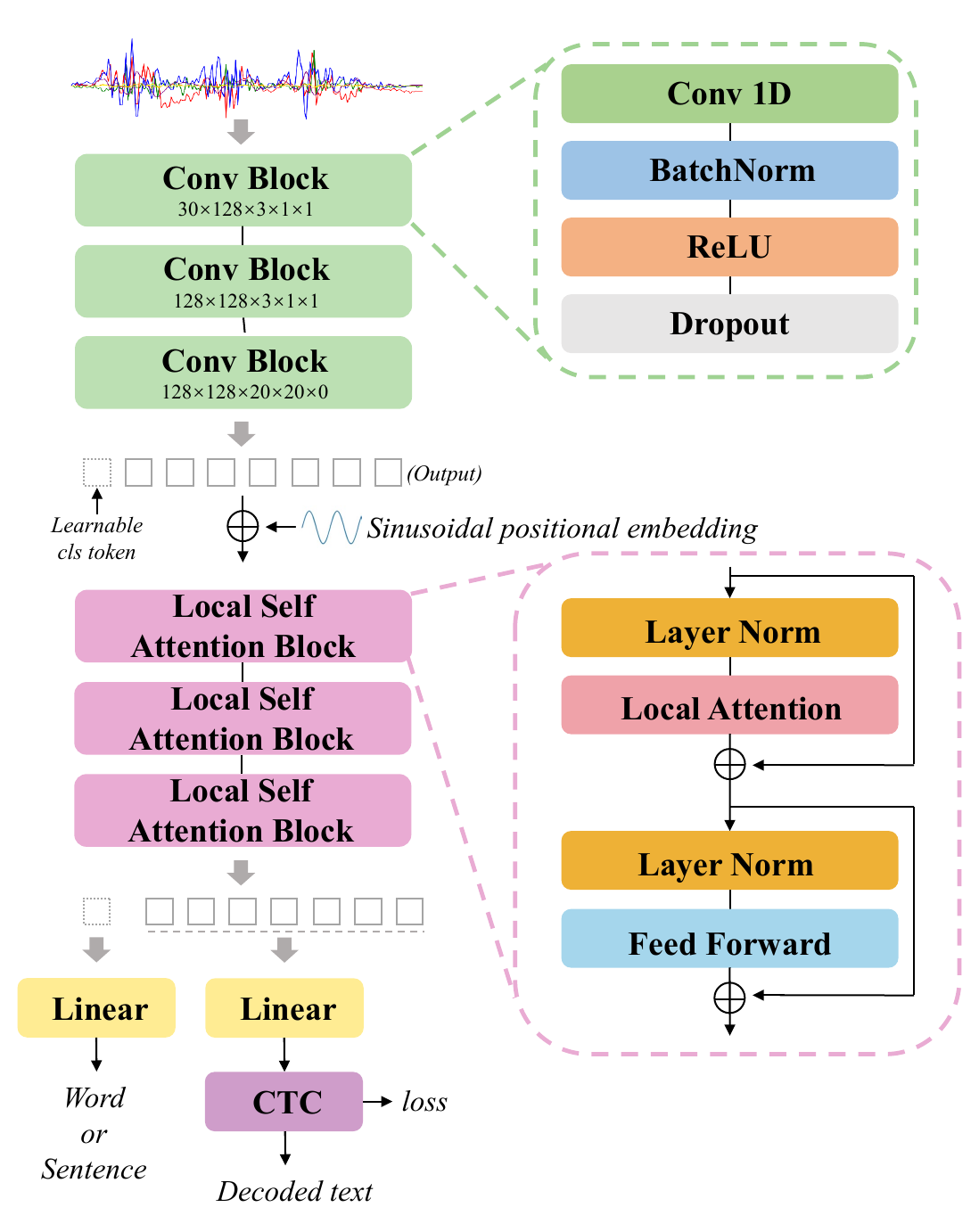}}
\caption{Architecture and operating principle of the network. Architecture of the network with three convolutional blocks, three local self-attention blocks, and linear layers.}
\label{fig:network}
\end{figure}

Fig.\ref{fig:network} shows the complete model structure. The model mainly consists of three parts: A convolution head that processes the accelerometer data by extracting local patterns and down-sampling the data; local self-attention blocks that aggregate the context information; and the CTC algorithm that decodes the features extracted from the previous network.

Conformer is a stacked architecture combining convolutional layers and Transformer blocks. Input features pass through three convolutional blocks, each consisting of a convolution layer, batch normalization, ReLU activation, and dropout. The first block maps 30 input channels to 128 hidden dimensions. The second incorporates a residual connection, and the third performs down-sampling to improve computational efficiency.

Local self-attention restricts the attention to a fixed window around each token, forming a band matrix. To preserve global context, a global token is prepended to each sequence to aggregate overall information.

The Connectionist Temporal Classification (CTC) algorithm performs well in sequence-to-sequence tasks. It enables the model to ignore irrelevant input frames and map multiple frames to a sequence of outputs. Its loss function is expressed as:

\begin{equation}
L_{CTC} = -\log{\sum_{\pi \in \Pi}P(\pi|X)}
\label{eq: CTCLoss}
\end{equation}

Where $\Pi$ is the set of all possible alignments of the target sequence $Y$ with the input sequence $X$, and $P(\pi|X)$ is the probability of alignment $\pi$ given the input sequence $X$. The probability of an alignment $\pi$ can be calculated by:

\begin{equation}
P(\pi|X) = \prod_{t=1}^T P(\pi_t|X_t)
\end{equation}

Where $\pi_t$ is the label at time step $t$ in the alignment $\pi$, and $P(\pi_t|X_t)$ is the probability of label $\pi_t$ at time step $t$ given the input $X_t$.

\subsubsection{Training Parameters}

The model was trained to minimize the sum of CTC loss and Cross-Entropy loss, using the Adams optimizer. The learning rate was set to 0.001, with a decay rate of 0.00001.  The batch size was 32 and the number of epochs was set to 50. PyTorch is used for all data training processes of the neural network mentioned above.

\section{Results}
\label{sec:results}

\subsection{Accuracy of Recognition}


The model shows an excellent performance in sentence segmentation and recognition. Accuracy is calculated as the ratio of correctly recognized words to the total words in a sentence or dataset. The model achieves an accuracy of 97.17\%, outperforming the results of 94.65 ± 2.54\% reported in previous studies \cite{kwon2023novel}. It attains an average classification accuracy of 97.24\% on words and 97.16\% on sentences.


\begin{figure}[t]

\begin{minipage}[b]{1.0\linewidth}
  \centering
  \centerline{\includegraphics[width=8.5cm]{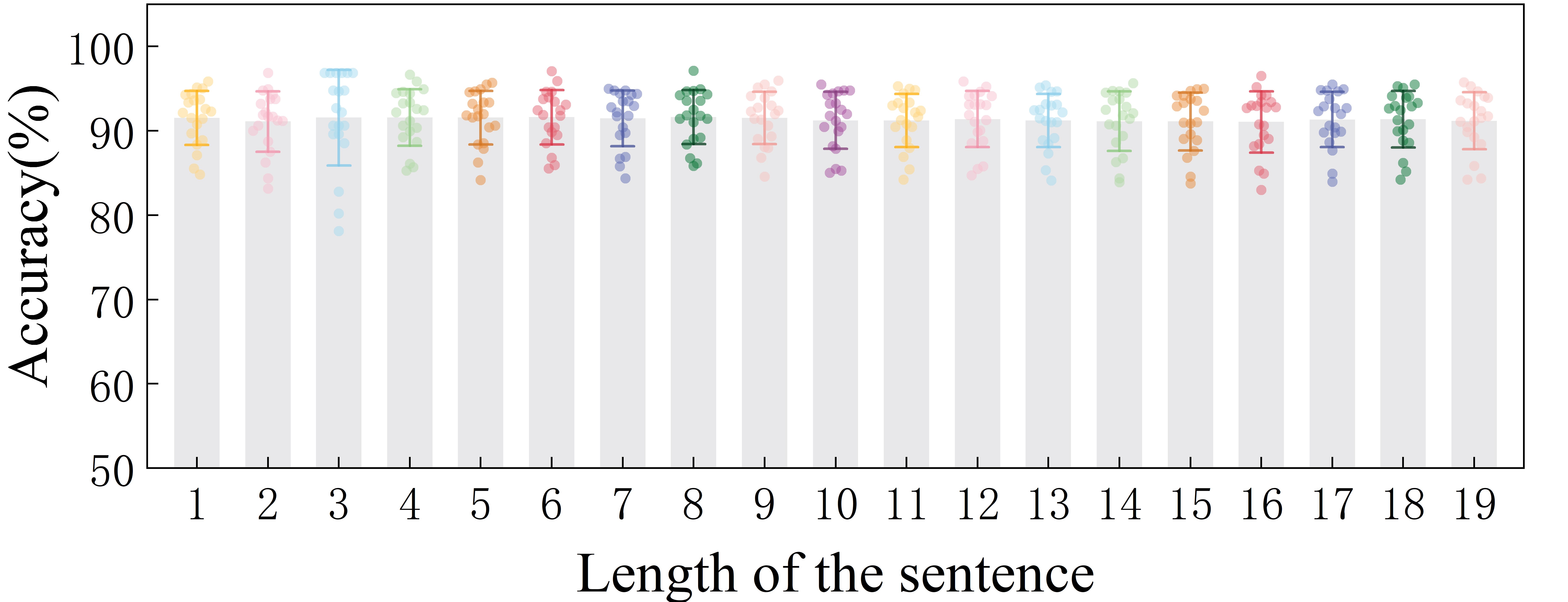}}
  \centerline{(a)}\medskip
\end{minipage}
%
\begin{minipage}[b]{1\linewidth}
  \centering
  \centerline{\includegraphics[width=8.5cm]{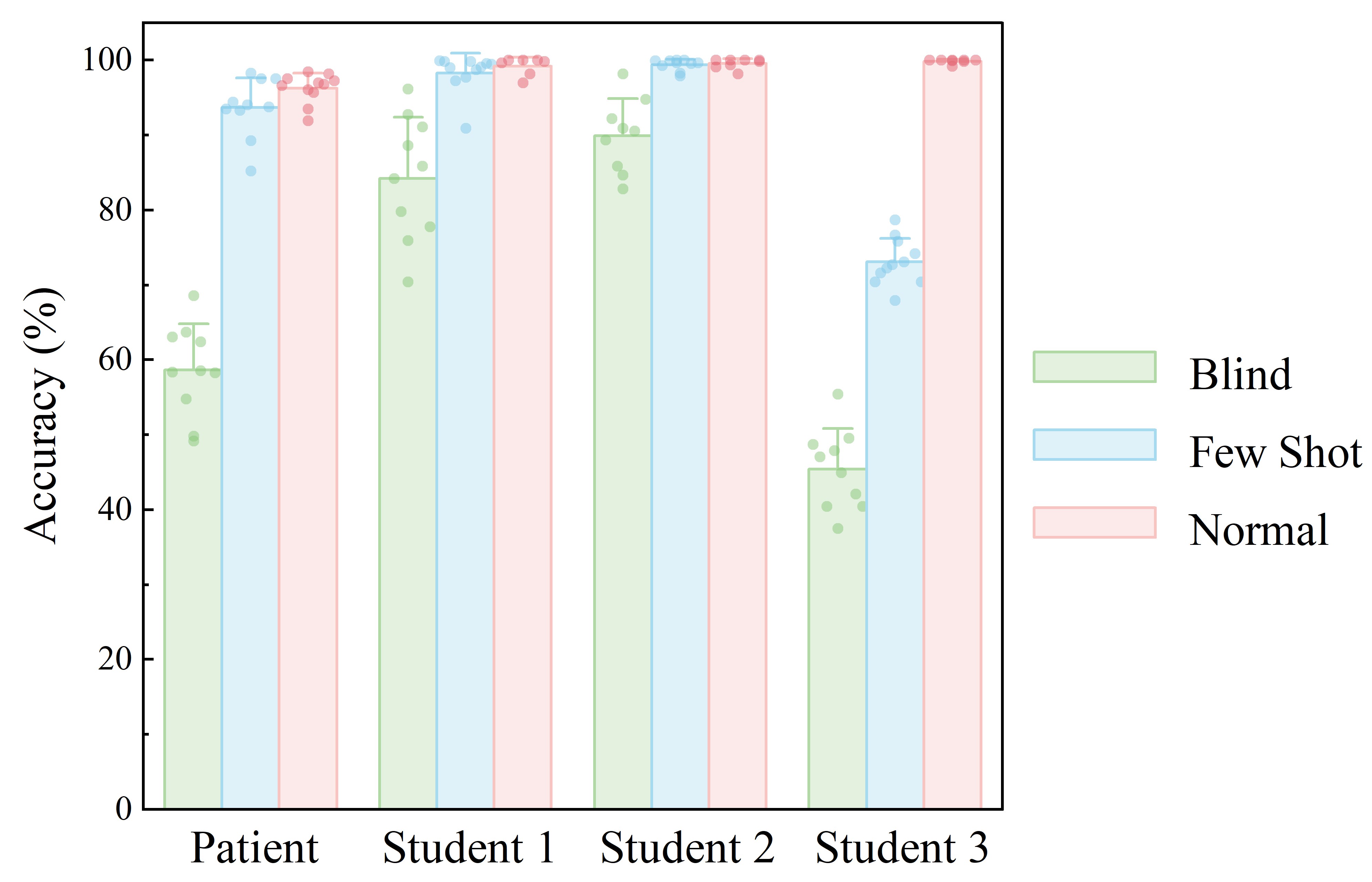}}
  \centerline{(b)}\medskip
\end{minipage}
\caption{Recognition related performance. (a) Histogram of the accuracy related to the length of the sentences; error bars indicate standard deviations. The accuracy remains above 90\% even when the length increases. (b) Histogram of the accuracy related to different participants. The error bars indicate standard deviations.}
\label{fig:segment}
\end{figure}

\subsection{Performance of Sentence Recognition}

The average sentence recognition accuracy reaches 97.16\%, indicating a high level of reliability in translating facial motion patterns into corresponding textual representations.

Fig.\ref{fig:segment}(a) presents the relationship between the average recognition accuracy and sentence length. The results show a relatively stable trend, with accuracy consistently remaining above 90\% across different sentence lengths. This demonstrates that unlike traditional methods, the CTC decoding effectively mitigates the impact of sentence length on recognition performance.


\subsection{Performance with respect to Different Individuals}

The model’s performance across different participants is illustrated in Fig.\ref{fig:segment}(b). The red bars in the figure represent the accuracy achieved by each participant under standard conditions, while the green bars depict “blind” testing results. In the “blind” scenario, the model was trained across different participants to assess the generalization capability of the model when applied to unseen subjects.

The results indicate that, on average, over 95\% accuracy is achieved in standard condition, while the “blind” accuracy varied among individuals, reflecting differences in speaking habits, facial muscle movements, and pronunciation styles.

To enhance the model’s generalization ability, a “few-shot” learning approach was applied. Incorporating a small number of participant-specific samples significantly improved recognition accuracy. The blue bars in Fig.\ref{fig:segment}(b) indicate that adding only a limited amount of personalized data, the model’s accuracy increased to over 90\%, effectively reducing the performance gap observed in blind testing.

\begin{figure}[t]
\centering
\centerline{\includegraphics[width=8.5cm]{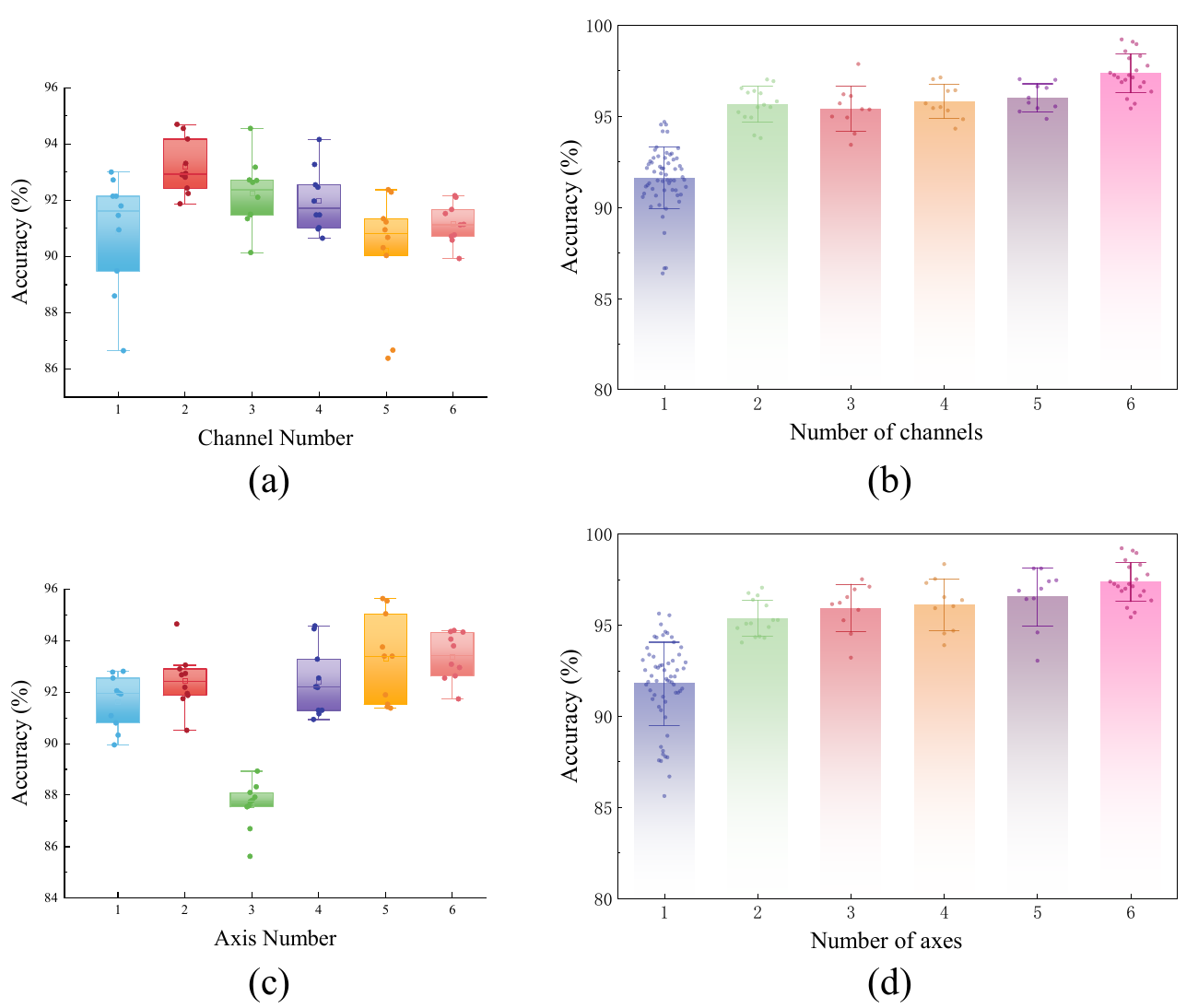}}
\caption{Performance considering different channels and axes. (a) The accuracy of a single channel. (b) The accuracy with respect to the number of channels. (c) The accuracy of a single axis. (d) The accuracy with respect to the number of axes.}
\label{fig:results}
\end{figure}

\subsection{Performance across Channels and Axes}

The results reported above are based on six channels. To test potential simplification of the system, channels and axes are evaluated seperately.

The mean accuracy of single channel ranges from 90.22\% to 93.19\%, as shown in Fig.\ref{fig:results}(a).

Fig.\ref{fig:results}(b) shows the accuracy using different number of channels. The mean accuracy of two or three channels can maintain above 95\%, with only a 1.5\% reduction compared to six channels. This suggests the system can be simplified without significant performance loss.


In terms of sensor axes, the model was trained using three accelerometer (axes 1–3) and three gyroscope axes (axes 4–6). Single-axis evaluation (Fig.\ref{fig:results}(c)) showed that gyroscope data (92.39–93.39\%) consistently outperformed accelerometer data (87.62–92.42\%), likely due to the rotational nature of articulatory movements (e.g., jaw rotation around the temporomandibular joint) during speech.

When increasing the number of axes, using two axes achieved 95.37\% accuracy, comparable to the 95.57\% attained with all six, as shown in Fig.\ref{fig:results}(d).


\section{Discussion}
\label{sec:discussion}

The strength of the proposed method lies in its ability to dynamically recognize arbitrary silent sentences by autonomous segmentation and identification. This overcomes the limitations of previous studies that treated word classification or treated sentences as a whole entirety \cite{wang2012sentence, luo2021end}. The previous methods are only capable of static recognition and lack flexibility and expressiveness, while this approach allows for a more versatile and practical SSI system.

Further work is required to improve its generalization capability. The relatively small dataset used in this study limits the model’s ability to handle complex linguistic phenomena. Expanding the database with diverse sentences and linguistic features could further enhance robustness and accuracy performance. Furthermore, instead of just the recognition of word sequences, the direct synthesis of speech can be examined. Based on the direct learning of speech by the accelerometer, we hope this system can enable generative speech communication in the future.

In addition to model performance, the practical deployment of silent speech interfaces must consider sensor comfort. Future iterations could explore more ergonomic designs, such as thinner flexible substrates or skin-conformal electronics, to enhance wearability over extended periods.

The proposed method offers a promising solution for individuals with speech and communication impairments, paving the way for more practical SSI systems. Future research would focus on addressing the challenges mentioned above and exploring to achieve better accuracy and generalization.

\section{Conclusion}
\label{sec:conclusion}

This paper presents a silent sentence recognition system using accelerometers and a Conformer-based network to capture facial motions and translate them into word and sentence transcriptions. The model achieves an average accuracy of 97.17\% at both word and sentence levels, demonstrating strong performance on classification and generalization.

As the first dynamic silent sentence recognition approach, the system offers portability and practical usability for patients in daily scenarios. Future work will explore multi-modal integration with microphones and other sensors.




\vfill\pagebreak

\section{Acknowledgement}

The authors would like to thank Cancer Hospital Chinese
Academy of Medical Sciences and volunteering participants
for help in obtaining the dataset. We also thank Professor
Heng-Ming Tai for his help on refining our manuscript.

\bibliographystyle{IEEEbib}
\bibliography{refs}

\end{document}